\documentclass[a4paper,twocolumn,superbib,nofootinbib,showkeys,superscriptaddress,amsmath,amssymb]{revtex4}

\usepackage{color,graphicx}
\usepackage{bm}
\usepackage{url}
\usepackage{subfigure}
\begin{document}

\title{Projective Synchronization and Control of Unified Chaotic System}
\thanks{Corresponding author: genle\_npu@hotmail.com}
\author{Liang Feng}
\affiliation{Department of Marine, Northwestern Polytechnical University, Xi'an, Shaanxi 710072, China}
\author{Xiang Jinglin }
\author{Chen Shaohua}
\author{Shi Jie}
\begin{abstract}
  The problem of projective synchronization(ps) and control are studied in modified
  unified chaotic system which possess partially linearity property. The desired ratio
  factor of corresponding subsystem variable could be obtained by state feedback control.
 Theoretical analysis and numerical simulations are provided to illustrate the projective
  synchronization and the feasibility of the proposed control method. The effect on projective
  synchronization caused by channel noise and parameter mismatch are investigated in detail,
  the results showed that parameter mismatch has more effect on projective synchronization than
  channel noise does, which may be applied to chaotic secure communications.
\end{abstract}

\keywords{unified chaotic system, projective synchronization, parameter mismatch}
\maketitle
\mdseries

\section{Introduction}
Since the seminal work by Pecora and Carroll\cite{Pecora1990}, the
synchronization of chaotic dynamical systems has been intensively
studied. Different types of synchronization phenomena have been
numerically observed and experimentally verified in a variety of
chaotic systems\cite{Pecora1997}. In 1999, Mainieri et al observed a
new phenomena in partially linear chaotic system, which they called
\emph{projective synchronization(ps)} \rule[1mm]{3.2mm}{0.2mm} the drive and response
vectors synchronize up to a scaling factor. This special synchronization concept
attracted increasing attention in past few years.

  Xu et al. investigated the stability criterion for projective synchronization in three
  dimensional chaotic systems\cite{Xu01a}, put forward a criteria for the occurrence of projective
  synchronization in chaotic systems of arbitrary dimension\cite{Xu01b}, and a necessary condition of
  projective synchronization in discrete-time systems of arbitrary dimensions\cite{Xu04a},
  and applied the projective synchronization technique to chaotic secure communication field\cite{Xu04b,Xu05}.
   Wang et al. studied the control of projective synchronization and put it into
    chaotic encryption application for the first time\cite{Wang04}.

   In 2004, Lu et al. presented a new unified chaotic system with continuous periodic switch(MLCL)which
 contains the Lorenz and Chen systems as two extremes and the L\"{u} system as a special case
in\cite{lu2004}, which has more complex nonlinear behavior and some properties, such as symmetry,
  invariance, dissipative and tracking any target function etc.

  Motivated by the aforementioned chaotic system, we investigate the projective synchronization
   and related aspects in MLCL system in detail. The rest of this paper is organized as follows. In the next section,
  we theoretically analyze the projective synchronization in MLCL chaotic system. In section III,
  a control scheme based state feedback are provided to obtain desired projective synchronization
  along with numerical simulations. The effect on projective synchronization caused by channel noise
   and parameter mismatch are investigated in section IV and V, respectively.
  Text is close with some concluding remarks.

\section{Projective synchronization in Chaotic system}
\vspace{-2mm}
\subsection{MLCL chaotic system}
\begin{figure}
   \centering
   \mbox{\subfigure[\hspace{1mm}$\omega=1.0$]  {\includegraphics[scale=0.31]{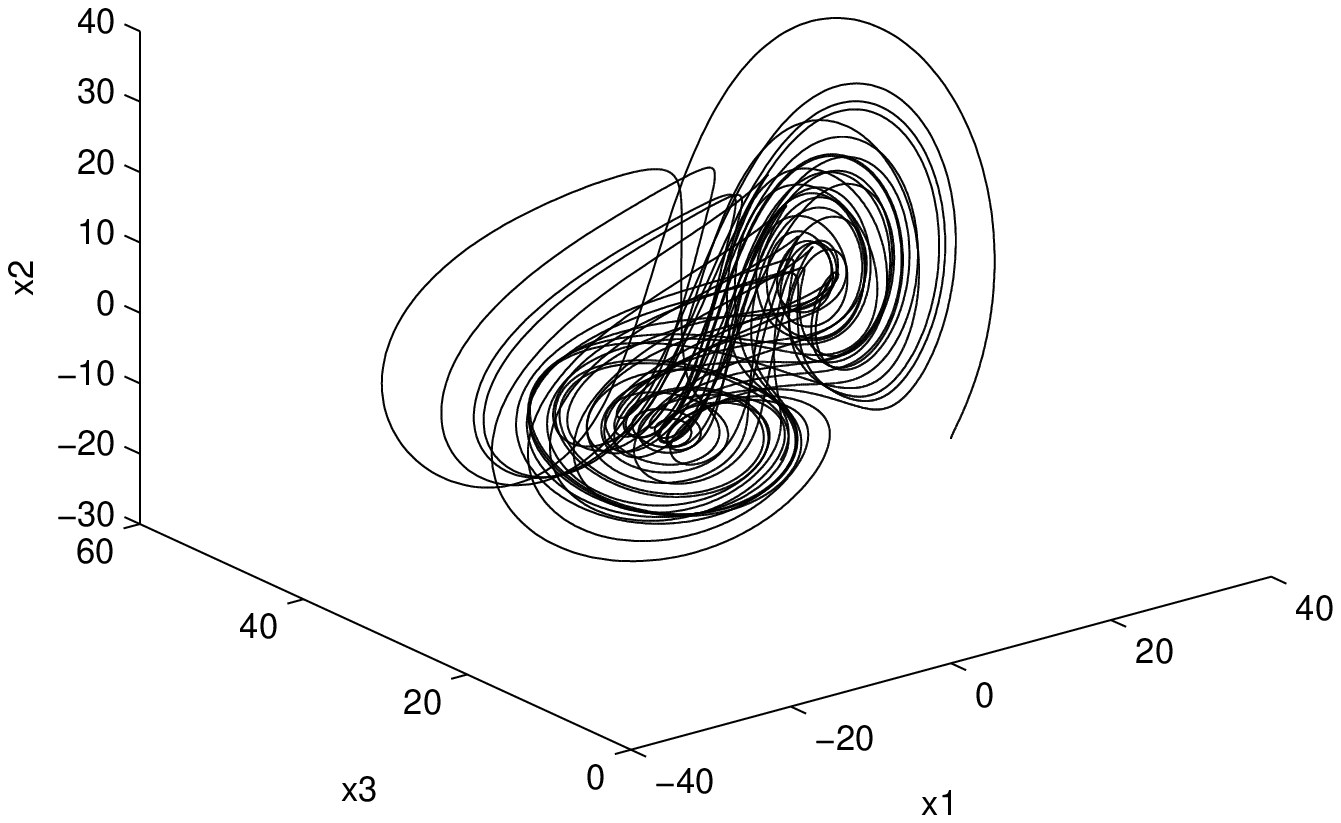}}\quad
         \subfigure[\hspace{1mm}$\omega=50$]  {\includegraphics[scale=0.31]{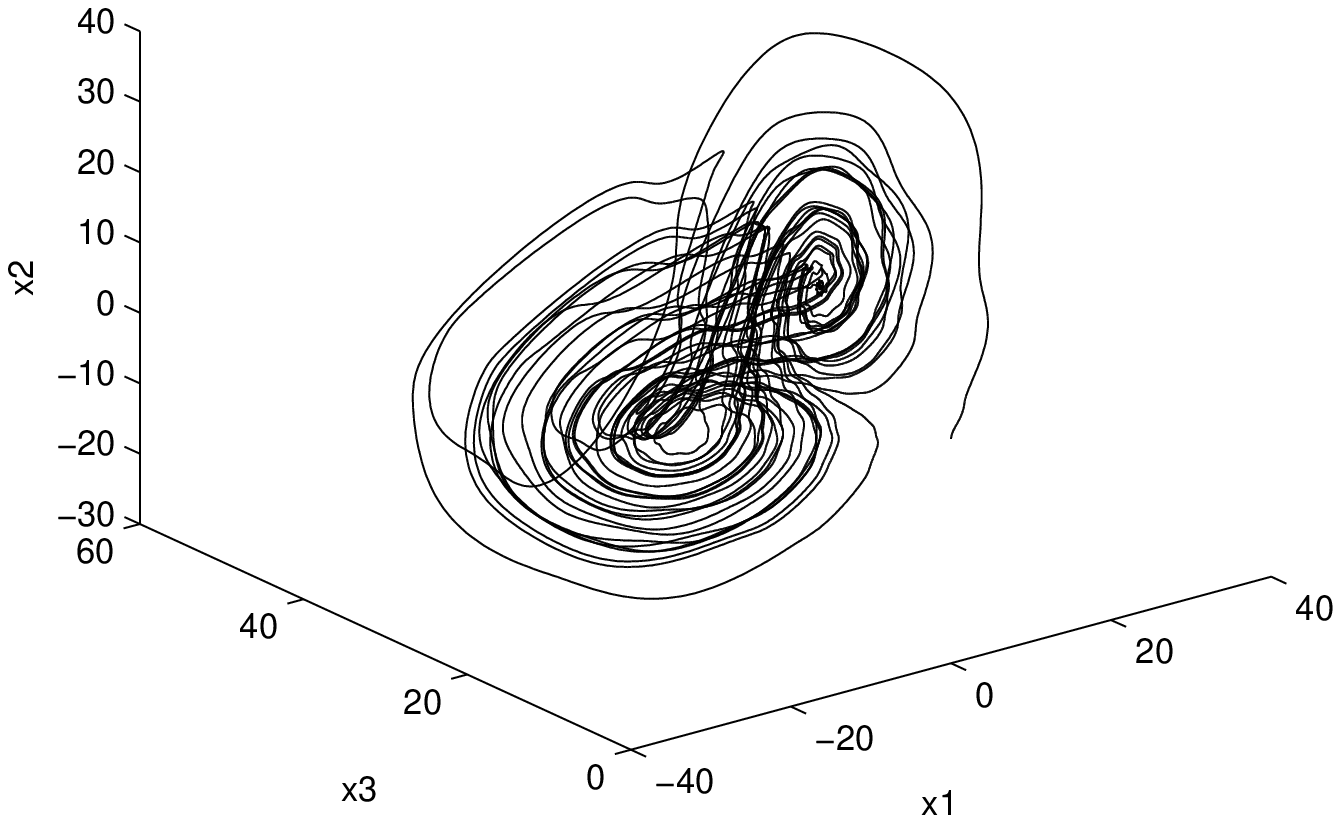}}}
  \vspace{-2mm}
  \caption{\small \textbf{System phase portrait for different $\omega$}}
  \label{fig:lclswitch}
\end{figure}

\vspace{-2mm}
   Lorenz had found the first classical chaotic attractor\cite{Lorenz1963} in 1963,
 Chen and Ueta have found a similar but non-equivalent attractor in 1999, which is known
 to be the dual of the Lorenz system. Recently, a chaotic system is presented by L\"{u} et al.,
  which bridged the gap between the Lorenz and Chen systems\cite{lu2002}.
  And a new unified chaotic system with continuous periodic switch(MLCL) between the Lorenz and
  Chen system is presented in \cite{lu2004} under inspiration of \cite{lu2002}. The MLCL chaotic
   system is described by:
\begin{equation}\label{equ:Unitedswitch}
\begin{cases}
    \dot{x}_{1}=(25a)+10)(x_{2}-x_{1})\\
    \dot{x}_{2}=(28-35a)x_{1}-x_{1}x_{3}+(29a-1)x_{2}\\
    \dot{x}_{3}=x_{3}x_{2}-\frac{8+a}{3}x_{3}
\end{cases}
\end{equation}
   where $a=\sin^{2}(\omega t)$ and $\omega$ is an adjustable parameter. System (\ref{equ:Unitedswitch})
   is a non-autonomous system, with the increase of t, system switches continuously between the Lorenz
   and Chen system, and the switching frequency is determined by the parameter $\omega$. The largest Lyapunov
    exponent (LLE) of system (\ref{equ:Unitedswitch}) increases with parameter $\omega$ increasing.
   In the following all the differential equations are solved using fourth-order Runge-Kutta
   in Mtalab. The abundant dynamics for different value of parameter $\omega$ are displayed in
   Fig.\ref{fig:lclswitch} under initial condition $x(0)=[0.01,0.01,0.01]$.
\vspace{-2mm}
\subsection{Projective synchronization in MLCL system}
\vspace{-2mm}
    System Eqs.(\ref{equ:Unitedswitch}) can be rewrite as follow:
\begin{equation}
\begin{cases}
   \dot{\mathbf{x}} =M(x_{3})\mathbf{x} \\
    \dot{x}_{3}=g(\mathbf{x},x_{3})=x_{1}x_{2}-\frac{8+a}{3}x_{3}
\end{cases}
\end{equation}
where $M(x_{3})=\begin{bmatrix}-25a-10) & 25a+10\\28-35a-x_{3} & 29a-1\end{bmatrix}$,
$\mathbf{x}=(x_{1},x_{2})^{\small T}$. The two systems are coupled through $z$, the $z$ in the response system
will be the $z$ of the drive system. The resulting system is a set of five differential equations:
\begin{equation}\label{equ:Unitedcoupled}
\begin{cases}
    \dot{x}_{1m}=(25a+10)(x_{2m}-x_{1m})\\
    \dot{x}_{2m}=(28-35a)x_{1m}-x_{1m}x_{3m}+(29a-1)x_{2m}\\
    \dot{x}_{3m}=x_{1m}x_{2m}-\frac{8+a}{3}x_{3m}\\
    \dot{x}_{1s}=(25a+10)(x_{2s}-x_{1s})\\
    \dot{x}_{2s}=(28-35a)x_{1s}-x_{1s}x_{3m}+(29a-1)x_{2s}
\end{cases}
\end{equation}
where subscript $m$ denotes the drive part and $s$ the response part.
Let $\frac{x_{s}}{x_{m}}=\frac{x_{1s}}{x_{1m}}=\frac{x_{2s}}{x_{2m}}=\alpha$, where $\alpha$ is
the ratio factor of state variable amplitude. we define error function as, $e_{ps}=x_{1m}x_{2s}-x_{1s}x_{2m}$,
and construct the Lyapunov function $V(x)=\frac{1}{2}e_{ps}^{2}$. The time derivative of $V(x)$ along
system (\ref{equ:Unitedcoupled}) is
\begin{equation}\label{equ:lya}
 \begin{split}
   \dot{V}=e_{ps}\dot{e}_{ps} &=e[\dot{x}_{1m}x_{2s}+x_{1m}\dot{x}_{2s}-\dot{x}_{1s}x_{2m}-x_{1s}\dot{x}_{2m}]\\
   &=(4\sin^{2}(\omega t)-11)e_{ps}^{2}\\
   &=(8\sin^{2}(\omega t)-22)V
\end{split}
\end{equation}
Since $sin^{2}(t)\in[0,1]$, $8\sin^{2}(\omega t)-22<0$, then we have,\\
$ V(t)=V(0)exp[(8\sin^{2}(\omega t)-22)t]\rightarrow0$\\
 Therefore, the following equation is obtained:
\begin{equation}\label{euq:limit}
\lim_{t\rightarrow\infty }e=\lim_{t\rightarrow\infty }(x_{1m}x_{2s}-x_{1s}x_{2m})=0
\end{equation}
Error function asymptotically converges to zero, which means the projective synchronization occurs.

    In order to explain this clearly, we study the evolution of the angular phase and ratio
  factor $\alpha$in the cylindrical coordinates. Firstly,
$x_{1}=r\cos\phi,x_{2}=r\sin\phi,x_{3}=x_{3}$, from Eqs.(\ref{euq:limit}), we have,
$\underset{t\rightarrow\infty }\lim r_{s}r_{m}(cos\phi_{m}\sin\phi_{s}-\cos\phi_{s}\sin\phi_{m})=r_{s}r_{m}\sin(\phi_{s}-\phi_{m})=0$.
Since $r_{s},r_{m}$ could not be zero all the time(if $r_{s}=0$, the stabilized phenomena
occurs explain in following context. if $r_{m}$ becomes nearly zero, some jumping points
will occur along with $\alpha$ evolution, both cases have no substantial impact upon subsequent
analysis and conclusion), so we have, $\underset{t\rightarrow\infty}\lim\sin(\phi_{s}-\phi_{m})=0$.
It shows that frequency locking phenomena occurs in coupled system after some time of evolution.

Secondly, The time derivative of $\alpha$ along system (\ref{equ:Unitedcoupled}) is:
\begin{equation}\label{euq:limitalpha}
 \begin{split}
  \dot{\alpha}& =\alpha(\frac{\dot{r}_{s}}{r_{s}}-\frac{\dot{r}_{m}}{r_{m}})\\
            &=\alpha[(54\sin^{2}(\omega t)-9)\sin(\phi_{m}+\phi_{s})+(38-10\sin^{2}(\omega t)\\
            &  -x_{3m})\cos(\phi_{m}+\phi_{s})]\sin(\phi_{s}-\phi_{m})\\
            &=\alpha*func(x_{3m},\phi_{m},\phi_{s})\sin(\phi_{s}-\phi_{m})
  \end{split}
\end{equation}
since $\lim_{t\rightarrow\infty }\sin(\phi_{s}-\phi_{m})=0$, then we have,
$\underset{t\rightarrow\infty}\lim \dot{\alpha}=0$, which means that factor $\alpha$
converges to a constant after some time of evolution.

\begin{figure}[!htb]
   \centering
   \includegraphics[scale=0.5]{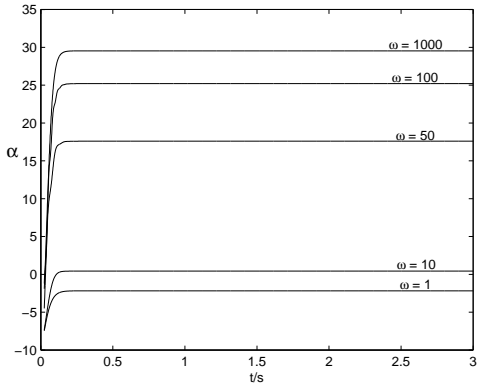}
   \vspace{-2mm}
   \caption{\small \textbf{Factor $\alpha$ under different $\omega$}}
   \label{fig:lclfactor}
\end{figure}

   Fig.\ref{fig:lclfactor} denotes the ratio factor $\alpha$ for different value
 of parameter $\omega$ under initial condition $[{\mathbf{x}}_{m}(0),{\mathbf{x}}_{s}(0)]=[0.1,0-0.1,-1,1]$.

It showed that the parameter $\omega$ has little effect on factor $\alpha$ provided $\omega<1$.
 when $\omega>1$, $\alpha$ increases with parameter $\omega$ increasing. when $\omega_{c} \simeq 8.935$,
 $\alpha$ nearly becomes zero (which means the response system was equivalently
 stabilized to origin, and the value $\omega_{c}$ is greatly dependent on initial condition),
 and the phase changes from negative to positive after $\alpha$
 increasing continuously. In the whole process, the projective synchronization
 can achieve up to $10^{-6}\thicksim10^{-8}$ precision level regardless of the transient time,
 and higher $\omega$ higher precision. Here the precision is defined as:
\begin{equation}\label{euq:limitalpha}
 p=\frac{std(\alpha-mean(\alpha))}{mean(\alpha)}
\end{equation}
where $std$ denotes the standard deviation and $mean$ the average.
But the value of $\alpha$ could not be predicted. The varying tendency of factor
$\alpha$ is well consistent with the results in literature\cite{liujie03}.

In addition, an interest phenomena is found, if the initial condition happens to satisfy the condition,
 $\frac{x_{1s}(0)}{x_{1m}(0)}=\frac{x_{2s}(0)}{x_{2m}(0)}=\alpha_{0}$, then factor $\alpha$
 will keep unchanged during the evolution.

  when parameter $\omega$ is fixed at $20$, factor $\alpha$ varies totally randomly under
 different initial conditions, other than the prediction of value of $\alpha$. The above results
  accord with the fact that chaotic systems are highly sensitive to initial condition and system parameters.

\section{Control on the projective synchronization}

     In this section, we presented a continuous control method to obtain desired factor
 $\alpha_{\small d}$ based on system states feedback technique under initial condition of
 $\omega=20$ and $[{\mathbf{x}}_{m}(0),{\mathbf{x}}_{s}(0)]=[0.1,0-0.1,-1,1]$.
 Two schemes, linear feedback and nonlinear feedback, are taken into consideration.

\begin{figure}[!htb]
   \centering
   \includegraphics[scale=0.5]{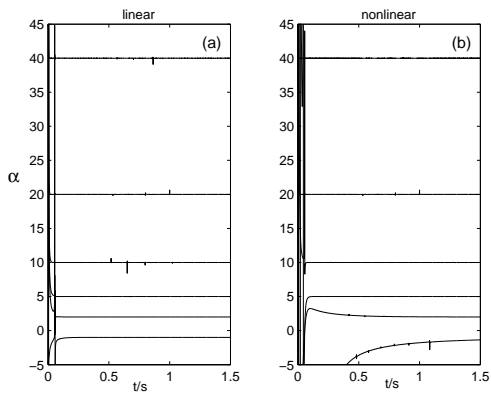}
   \vspace{-3mm}
   \caption{\small Projective syn. curve based on feedback control}
   \label{fig:lclpscon}
\end{figure}

 In case of linear feedback, let control item $U=k(x_{s}-\alpha_{\small d}x_{m})$,
 where $k$ denotes the strength of feedback, adding $U$ to the drive part of coupled
  system(\ref{equ:Unitedcoupled})(if adding $U$ to response part, similar results can also be obtained):
\begin{equation}\label{equ:lclpscoupledcon}
\begin{cases}
    \dot{x}_{1m}=(25a+10)(x_{2m}-x_{1m}) + k(x_{1s}-\alpha_{\small d}x_{1m}) \\
    \dot{x}_{2m}=(28-35a)x_{1m}-x_{1m}x_{3m}+(29a-1)x_{2m} \\
    \hspace{15mm}+ k(x_{2s}-\alpha_{\small d}x_{2m}) \\
    \dot{x}_{3m}=x_{1m}x_{2m}-\frac{8+a}{3}x_{3m}\\
    \dot{x}_{1s}=(25a+10)(x_{2s}-x_{1s}) \\
    \dot{x}_{2s}=(28-35a)x_{1s}-x_{1s}x_{3m}+(29a-1)x_{2s}
\end{cases}
\end{equation}
The projective synchronization curve is displayed in Fig.\ref{fig:lclpscon}(a).

  In case of nonlinear feedback, $U=\alpha_{\small d}(x_{s}-\alpha_{\small d}x_{m})$,
  and other parameters keep the same, the projective synchronization curve
  is plotted in Fig.\ref{fig:lclpscon}(b).

   From Fig.\ref{fig:lclpscon} and much simulation, we found:
  \begin{enumerate}
    \item Projective synchronization occurs in short time and desired ratio factor
    $\alpha_{\small d}$ is obtained using both methods.

    \item In comparison, under small value of $\alpha$, linear feedback method has faster
     control response speed than nonlinear feedback does, while under large value of $\alpha$,
     the control response speed is comparative.

    \item The synchronization precision both degrade slightly compared with the uncontrolled
     coupled system (\ref{equ:Unitedcoupled}) and wave error occurs, but the precision
     is still able to achieve $10^{-3}\thicksim10^{-4}$ level, and larger value of
     $\alpha$, higher precision.

    \item Large value of $\alpha$ could be obtained successfully by both methods, while such large
     $\alpha$ can not be observed in uncontrolled coupled system (\ref{equ:Unitedcoupled}).
     In order to add evidence, we have plotted the drive and response subsystem phase portrait
      for ratio factor $\alpha=3$. 
  \end{enumerate}
\begin{figure}[!htb]
   \centering
   \includegraphics[scale=0.5]{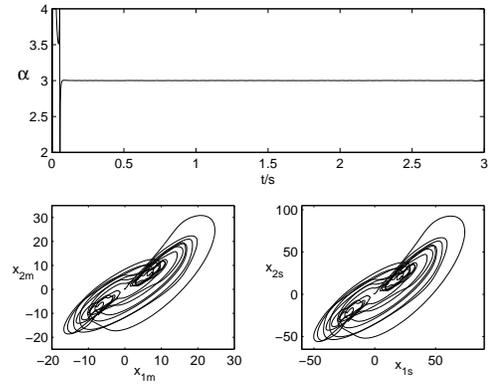}
   \vspace{-2mm}
   \caption{\small  space phase of coupled under $\alpha=3$}
   \label{fig:lclpsphase}
\end{figure}

\section{ Effect on PS caused by channel noise and parameter mismatch}

      In practical scenario, channel noise and system parameter mismatch are unavoidable.
In this section, we investigate the effect on ps caused by such imperfection.
 the following figures are based on condition of $k=20, \omega=20$.

\begin{figure}[!htb]
   \centering
   \includegraphics[scale=0.5]{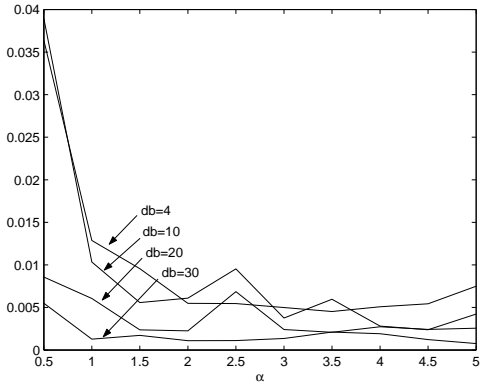}
   \vspace{-3mm}
   \caption{\small Effect on ps caused by channel noise}
   \label{fig:psnoise}
\end{figure}

    From Fig.\ref{fig:psnoise} and much numerical simulation, we found that under
  certain SNR($SNR>=10dB$), channel noise has less effect on ps, the precision
  is still able to achieve $10^{-2}\thicksim10^{-3}$ level. PS could occur if taken
  some noise-reduction method. In addition, the effect from parameter $k$ mainly lies in
  control time $t_{ps}$, and $t_{ps}$ decreases while $k$ increases.
  The effect from parameter $\omega$ put little influence on ps partially owing
  to channel noise containing much frequency ingredient.

\begin{figure}[!htb]
   \centering
   \includegraphics[scale=0.5]{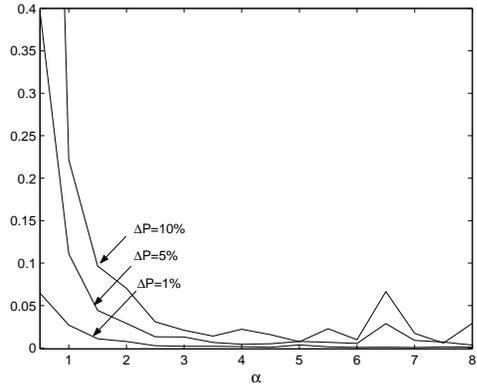}
   \vspace{-3mm}
   \caption{\small Effect on ps caused by parameter mismatch}
   \label{fig:psmis}
\end{figure}

    From Fig.\ref{fig:psmis}, we found that when parameter mismatch is limited to
    range of $[0,10\%]$, ps can be achieved successfully with $10^{-2}\thicksim10^{-3}$
    synchronization precision. The effect from parameter $k$ lies in the control time $t_{ps}$
    mentioned above. Parameter $\omega$ has more effect on ps compared with aforementioned case.

 In general, parameter mismatch has much influence on ps than channel noise does,
    ps could be achieved with certain precision provided taken some effective measure,
    this property can be applied to chaos control and chaotic secure communications.

\section{Conclusions}
\vspace{-2mm}
    In this paper, we have investigated the projective synchronization properties in MLCM
    chaotic system, and present control scheme based continuous state feedback to control
    the MLCL system to obtain desired ratio factor $\alpha$,
    The effectiveness and feasibility of our methods have been verified by
     computer simulation. The effect on projective synchronization caused by channel noise
     and parameter mismatch are investigated, the results showed that parameter mismatch has
     more effect on projective synchronization than channel noise does, which may be applied to
     chaotic secure communications.

      To our best knowledge, this is the first report on the projective synchronization
      of modified unified chaotic system. From the viewpoint of system energy, the small ratio
      factor are desired to obtained, and discrete impulsive control scheme are considered to
      replace the continuous feedback control counterpart.

\bibliographystyle{unsrt}


\end{document}